\documentclass[usenatbib]{mn2e}

\usepackage{savesym}
\usepackage{amsmath}
\savesymbol{iint}
\usepackage{txfonts}
\usepackage{deluxetable}
\usepackage{html}

\usepackage{subfigure}
\restoresymbol{TXF}{iint}

\usepackage{color}
\usepackage[normalem]{ulem}

\usepackage{subfigure}
\usepackage{graphicx}
\usepackage{epstopdf}
\usepackage{url}

\def\simless{\mathbin{\lower 3pt\hbox
{$\rlap{\raise 5pt\hbox{$\char'074$}}\mathchar"7218$}}}   
\def\simmore{\mathbin{\lower 3pt\hbox
{$\rlap{\raise 5pt\hbox{$\char'076$}}\mathchar"7218$}}}   
\newcommand{\be}{\begin{equation}}
\newcommand{\ee}{\end{equation}}
\newcommand       \bea          {\begin{eqnarray}}
\newcommand       \eea          {\end{eqnarray}}
\newcommand       \apj          {ApJ}
\newcommand       \apjl         {ApJL}
\newcommand       \aap          {A\&A}
\newcommand       \nat          {Nature}
\newcommand       \mnras        {MNRAS}
\newcommand       \pasp      {PASP}
\newcommand       \aj      {AJ}
\newcommand       \prd      {Phys.~Rev.~D.~}

\newcommand       \araa      {ARA\&A}

\newcommand      \apjs {ApJ Supplements}

\def\simlt{\mathrel{\hbox{\rlap{\hbox{\lower4pt\hbox{$\sim$}}}\hbox{$<$}}}}
\def\simgt{\mathrel{\hbox{\rlap{\hbox{\lower4pt\hbox{$\sim$}}}\hbox{$>$}}}}
\def\lesssim{\mathrel{\hbox{\rlap{\hbox{\lower4pt\hbox{$\sim$}}}\hbox{$<$}}}}

\def\gtrsim{\mathrel{\hbox{\rlap{\hbox{\lower4pt\hbox{$\sim$}}}\hbox{$>$}}}}

\title[Gamma-ray novae as probes of particle acceleration]{Gamma-ray novae as probes of relativistic particle acceleration at non-relativistic shocks}

\author[]{B.~D.~Metzger$^{1}\thanks{E-mail: bmetzger@phys.columbia.edu}$, T.~Finzell$^{2}$, I.~Vurm$^{1}$, R.~Hasco\"{e}t$^{1}$, A.~M.~Beloborodov$^{1}$, L.~Chomiuk$^{2}$\\
$^{1}$Columbia Astrophysics Laboratory, New York, NY, 10027, USA\\
$^{2}$Department of Physics and Astronomy, Michigan State University, East Lansing, MI 48824, USA\\
}

\begin{document}
\date{Received / Accepted}
\pagerange{\pageref{firstpage}--\pageref{lastpage}} \pubyear{2014}

\maketitle

\label{firstpage}

\begin{abstract}

The {\it Fermi} LAT discovery that classical novae produce $\gtrsim 100$ MeV gamma-rays establishes that shocks and relativistic particle acceleration are key features of these events.  These shocks are likely to be radiative due to the high densities of the nova ejecta at early times coincident with the gamma-ray emission.  Thermal X-rays radiated behind the shock are absorbed by neutral gas and reprocessed into optical emission, similar to Type IIn (interacting) supernovae.  Gamma-rays are produced by collisions between relativistic protons with the nova ejecta (hadronic scenario) or Inverse Compton/bremsstrahlung emission from relativistic electrons (leptonic scenario), where in both scenarios the efficiency for converting relativistic particle energy into LAT gamma-rays is at most a few tens of per cent.  The measured ratio of gamma-ray and optical luminosities, $L_{\gamma}/L_{\rm opt}$, thus sets a lower limit on the fraction of the shock power used to accelerate relativistic particles, $\epsilon_{\rm nth}$.  The measured value of $L_{\gamma}/L_{\rm opt}$ for two classical novae, V1324 Sco and V339 Del, constrains $\epsilon_{\rm nth} \gtrsim 10^{-2}$ and $\gtrsim 10^{-3}$, respectively.  Leptonic models for the gamma-ray emission are disfavored given the low electron acceleration efficiency, $\epsilon_{\rm nth} \sim 10^{-4}-10^{-3}$, inferred from observations of Galactic cosmic rays and particle-in-cell (PIC) numerical simulations. A fraction $f_{\rm sh} \gtrsim 100(\epsilon_{\rm nth}/0.01)^{-1}$ and $\gtrsim 10(\epsilon_{\rm nth}/0.01)^{-1}$ per cent of the optical luminosity is powered by shocks in nova Sco and nova Del, respectively.  Such high fractions challenge standard models that instead attribute all nova optical emission to the direct outwards transport of thermal energy released near the white dwarf surface.  We predict hard $\sim 10-100$ keV X-ray emission coincident with the LAT emission, which should be detectable by NuSTAR or ASTRO-H, even at times when softer $\lesssim 10$ keV emission is absorbed by neutral gas ahead of the shocks.    
\end{abstract} 
  

\section{Introduction} 
\label{sec:intro}

As ``guest stars" to our ancestors, novae have been observed since antiquity (\citealt{Zwicky36}).  Powered by runaway nuclear burning on the surface of white dwarfs (e.g.~\citealt{Starrfield+72}), nova eruptions are usually accompanied by the expulsion of matter outwards at high velocities $\gtrsim 10^{3}$ km s$^{-1}$ (\citealt{Shore13}, and references therein).  Most of the radiated energy in a nova outburst occurs at optical and ultraviolet wavelengths and is generally attributed to the outwards transport of thermal energy produced near the white dwarf surface (\citealt{Prialnik86}; \citealt{Prialnik&Kovetz95}; \citealt{Starrfield+00}).  Dynamical stellar evolution and hydrodynamic calculations are used to interpret basic features of nova light curves, such as how their rate of evolution depends on the white dwarf mass, central temperature, and accretion rate (e.g., \citealt{Yaron+05}; \citealt{Hillman+14}).  Despite notable successes, models of nova outbursts remain plagued by uncertainties such as the efficiency of convective mixing and the assumed prescription for mass loss.  This makes it challenging to accurately predict the quantity, velocity, and time history of matter ejected from the white dwarf surface.

The Large Area Telescope (LAT) on the {\it Fermi} satellite recently discovered that classical novae emit $\gtrsim 100$ MeV gamma-rays, occurring at times coincident to within a few days of the optical peak and lasting for several weeks (\citealt{Ackermann+14}).  Luminous continuum gamma-ray emission indicates the presence of strong shocks which accelerate particles to relativistic energies, producing gamma-rays via leptonic or hadronic processes (e.g., \citealt{Tatischeff&Hernanz07}).  Radio VLBI imaging by \citet{Chomiuk+14} recently detected compact, high brightness temperature knots from V959 Mon, confirming the presence of non-thermal shock emission in a gamma-ray producing nova.  

Evidence for shocks in novae has existed well before {\it Fermi} in the form of hard X-ray emission (e.g., \citealt{Mukai&Ishida01}; \citealt{Mukai+08}), early peaks in the radio emission which are inconsistent with thermal emission from freely expanding photo-ionized gas (e.g., \citealt{Taylor+87}; \citealt{Krauss+11}; \citealt{Weston+13}), and coronal line emission during the nebular phase requiring a harder source of ionizing radiation than expected from the cooling white dwarf (e.g., \citealt{Shields&Ferland78}).  However, since most of these signatures are observed on timescales of months or later after the eruption, shocks appear to have been relegated to a mere side feature of the main thermonuclear event.  The {\it Fermi} discovery of luminous shocks in coincidence with the optical peak unambiguously establishes their importance to the qualitative picture of nova eruptions.  The dearth of X-ray and radio shock signatures at times coincident with the gamma-rays is unsurprising in retrospect due to the high bound-free and free-free optical depths created, respectively, by large columns of neutral and ionized gas at early times (\citealt{Hillman+14}; \citealt{Metzger+14}).  

Shocks were unexpected in classical novae because the pre-eruption environment surrounding the white dwarf is occupied only by the low density wind of the main sequence companion star, requiring a different source of matter into which the nova outflow collides.  One physical picture, consistent with both optical (e.g., \citealt{Schaefer+14}) and radio imaging (e.g.,\citealt{Chomiuk+14}), and the evolution of optical spectral lines (e.g.,~\citealt{Ribeiro+13}; \citealt{Shore+13}), is that the thermonuclear runaway is first accompanied by a slow ejection of mass with a toroidal geometry, the shape of which may be influenced by the binary companion (e.g., \citealt{Livio+90}; \citealt{Lloyd+97}).  This slow outflow is then followed by a second ejection or a continuous wind (e.g., \citealt{Bath&Shaviv76}) with a higher velocity and more spherical geometry.  

\begin{figure}
\subfigure{
\includegraphics[width=0.5\textwidth]{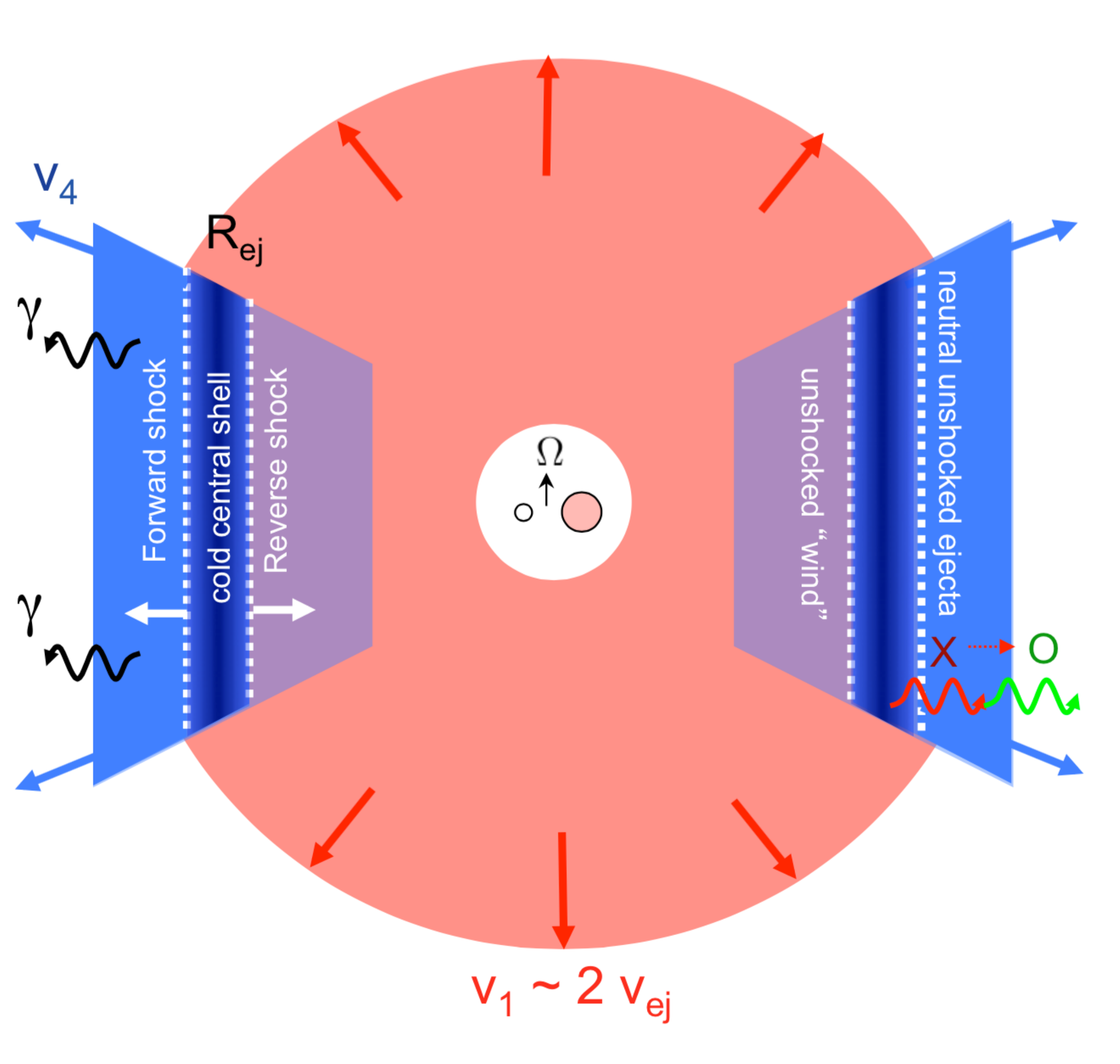}} 
\caption{Schematic diagram of the proposed geometry (side view) of shock interaction in classical nova outflows (see text).  A slow outflow with velocity $v_{4} \lesssim 10^{3}$ km s$^{-1}$ is ejected first, its geometry shaped into an equatorially-concentrated torus, possibly due to interaction with the binary companion of the white dwarf (e.g., ~\citealt{Livio+90}).  This outflow is followed within a few days by a faster outflow or continuous wind with a higher velocity $v_{1} \sim 2v_{4}$.  The fast and slow components collide, produce a forward-reverse shock structure.  Kinetic energy dissipated by the shocks is radiated as thermal X-rays, which are absorbed by neutral gas ahead or behind the shock and re-radiated as thermal optical/UV emission (\citealt{Metzger+14}).  A small fraction, $\epsilon_{\rm nth} \ll 1$, of the shock power is used to accelerate non-thermal ions or electrons, which radiate gamma-rays by interacting with ambient gas or radiation, respectively.}
\label{fig:schematic}
\end{figure}

Assuming the expanding ejecta cools sufficiently, the subsequent collision between the fast and slow components produces strong ``internal" shocks within the ejecta which are concentrated in the equatorial plane.  The fast component continues to expand freely along the polar direction, creating a bipolar morphology (Fig.~\ref{fig:schematic}).  Interestingly, modeling of the symbiotic gamma-ray novae V407 Cyg (\citealt{Abdo+10}) also appeared to require the presence of a dense equatorial torus (\citealt{Martin&Dubus13}), similar to that inferred in classical novae (see also \citealt{Sokoloski+08}, \citealt{Orlando+09}, \citealt{Drake+09}, \citealt{Orlando&Drake12} for other evidence for bipolar ejecta in symbiotic novae).    

\cite{Metzger+14} present a semi-analytic model for nova shocks and their thermal emission, which they fit to the radio and X-ray data of the gamma-ray nova V1324 Sco.   A key finding was that the shocks responsible for the radio maximum seen months after the eruption could, at smaller radii and earlier times, also power the nova optical emission.  One can generalize the basic argument as follows.  Gas ahead of a shock with power sufficient to create the observed gamma-ray emission is necessarily dense.  Most of the kinetic energy dissipated by shocks moving through a dense medium is radiated as thermal X-rays (a so-called ``radiative" shock).  These X-rays are absorbed by neutral gas ahead or behind the shock, reprocessing their energies to lower, optical frequencies, where the lower opacity readily allows their energy to escape.  The observed gamma-ray luminosity is typically only a fraction $\sim 10^{-4}-10^{-2}$ of that emitted as optical/UV radiation ($\S\ref{sec:data}$).  However, since only a fraction of the shock power is used to accelerate relativistic particles and only a fraction of that is radiated in the LAT bandpass, one concludes that {\it a significant fraction of the nova optical emission is shock-powered}.  

The above result also implies that combined optical and gamma-ray data from novae can be used to probe particle acceleration at non-relativistic shocks.  The basic concepts of diffusive shock acceleration (DSA) were developed almost forty years ago (e.g., \citealt{Axford+77}; \citealt{Bell78}; \citealt{Blandford&Ostriker78}).  However, only recently have plasma kinetic simulations seen the self-consistent development of the DSA cycle in collision-less non-relativistic shocks (e.g., \citealt{Caprioli&Spitkovsky14}; \citealt{Kato14}; \citealt{Park+14}).  As we shall discuss, gamma-ray novae provide a real-time laboratory for testing the predictions of DSA theory in ways complementary to traditional methods, such as the modeling of supernova remnant emission.

This paper solidifies the above theoretical arguments and puts them into practice using data from the classical novae V1324 Sco and V339 Del (hereafter, `nova Sco' and `nova Del', respectively).  Though well-studied at many wavelengths, we do not consider the nova V959 Mon because of the lack of early optical coverage concurrent with the gamma-ray detections.  The salient properties of nova shocks are reviewed in $\S\ref{sec:shocks}$, including their radiative nature ($\S\ref{sec:radiative}$), the high efficiency with which shock power is radiated at optical/UV frequencies ($\S\ref{sec:optical}$), and the efficiency of particle acceleration and gamma-ray production ($\S\ref{sec:shockacceleration}$).  A combined analysis of the optical and gamma-ray data of V1324 Sco and V3229 Del is presented in $\S\ref{sec:data}$, using our theoretical framework to constrain the minimum fraction of the shock power placed into relativistic particles and the minimum fraction of nova optical light curves that are shock powered.  In $\S\ref{sec:discussion}$ we discuss our results and their implications for particle acceleration at non-relativistic shocks and for the power source behind nova optical light curves.

\section{Shocks in Gamma-Ray Novae}
\label{sec:shocks}

\subsection{The shocks are probably radiative}
\label{sec:radiative}

The slow outflow from a nova with velocity $v_{4} = 10^{3}v_{8}$ km s$^{-1}$ expands to a radius
\be
R_{\rm ej} = v_{4} t \approx 6\times 10^{13}t_{\rm wk}v_{8}\,{\rm cm}
\ee
by a time $t = t_{\rm wk}$ week.  The density in the slow ejecta of assumed thickness $\sim R_{\rm ej}$ and hydrogen-dominated composition is given by
\be
n_{\rm ej} \approx \frac{M_{\rm ej}}{4\pi R_{\rm ej}^{3}f_{\Delta \Omega} m_p} \sim 9\times 10^{10}M_{-4}t_{\rm wk}^{-3}v_{8}^{-3}\,{\rm cm^{-3}},
\label{eq:nej}
\ee
where $f_{\Delta \Omega} \sim 0.5$ is the fraction of the total solid-angle subtended by the outflow (Fig.~\ref{fig:schematic}) and $M_{\rm ej} = 10^{-4}M_{-4}M_{\odot}$ is the ejecta mass, normalized to a value characteristic of those measured by late thermal radio emission (\citealt{Hjellming+79}; \citealt{Seaquist+80}).  We assume a radially-thick outflow because the resulting low density makes this the most conservative assumption when arguing for radiative shocks.  A wide solid angle $f_{\Delta \Omega} \sim 0.5$ is also the most conservative assumption when arguing for radiative shocks and is needed to produce a high efficiency for converting the kinetic energy of the outflow to shock, and hence gamma-ray, luminosity.

A faster outflow (hereafter ``wind") of mass loss rate $\dot{M}$ and velocity $v_{1} \sim 2 v_{4}$ collides with the ejecta from behind.  The relative velocity between the fast and slow outflows is unlikely to greatly exceed this assumed value $\sim v_{4}\sim  10^{3}$ km s$^{-1}$, but if it is much lower than assumed this would only strengthen our subsequent argument that the shocks are radiative, since the cooling time is an increasing function of the shock velocity.  The density of the wind at the collision radius ($\sim$ radius of the slow ejecta) is given by
\be
n_{\rm w} \approx \frac{\dot{M}}{4\pi R_{\rm ej}^{2}m_p v_{1}} \sim 2\times 10^{9}\dot{M}_{-5}v_{8}^{-3}t_{\rm wk}^{-2}{\,\rm cm^{-3}},
\label{eq:nw}
\ee
where $\dot{M} = 10^{-5}\dot{M}_{-5}M_{\odot}\,{\rm wk^{-1}}$ is normalized to a value resulting in the ejection of $\sim 10^{-5}M_{\odot}$ over a week.  For instance, a total mass $4\times 10^{-5}M_{\odot}$ was ejected in the ``fast" component of V959 Mon (\citealt{Chomiuk+14}).  Our assumption that the fast outflow is a steady wind is also the most conservative one when arguing for radiative shocks; for a fixed fast ejecta mass, the wind density at the reverse shock is smaller than if the ejection had occurred over a shorter duration.   

The interaction drives a forward shock (FS) through the slow shell and a reverse shock (RS) back through the wind (Fig.~\ref{fig:schematic}).  Assuming the shocks are radiative, the post shock material is compressed and piles up in a central cold shell sandwiched by (the ram pressure of) the two shocks.  The FS propagates at a velocity $v_{\rm f} = v_{\rm c}-v_{4} \ll v_{4}$, while the velocity of the reverse shock is $v_{\rm r} = v_{1}-v_{\rm c}$, where $v_{\rm c}$ is the velocity of the cold central shell (see \citealt{Metzger+14}, although be wary of notational\footnote{In particular, \citet{Metzger+14} define $v_{\rm sh}$ as the velocity of the central shell ($v_{\rm c}$ in this paper).  In this paper $v_{\rm sh}$ instead denotes the velocity of the shocks.} differences).  Here we have assumed the shocks are radiative, for which the velocity of the immediate post shock gas slows to that of the cold central shell, such that the shock velocity equals the difference between the velocity of the upstream gas and the central shell.  Hereafter, the velocities of both shocks are parametrized as $v_{\rm sh} = \eta v_{4}$, where we expect that $\eta < 1$ for the FS and $\eta \sim 1$ for RS.  Also note that for parameters of interest, the amount of kinetic energy dissipated by the RS is greater than that by the FS due to the higher velocity of the RS, potentially favoring the RS as the site of particle acceleration.

The shocks heat the gas to a temperature
\be
T_{\rm sh} \simeq \frac{3}{16 k}\mu m_p v_{\rm sh}^{2} \sim 1.4\times 10^{7}v_{8}^{2}\eta^{2}\,{\rm K}
\label{eq:Tsh}
\ee
and compresses it to a density $n_{\rm sh} = 4n_{\rm ej}$ (FS) or 4$n_{\rm w}$ (RS), where we have taken $\mu = 0.62$.  

Gas cools behind the shock on a characteristic timescale
\begin{eqnarray}
&& t_{\rm cool}   = \frac{3kT_{\rm sh}/2}{n_{\rm sh}\Lambda(T_{\rm sh})} \approx
\left\{
\begin{array}{lr}
 1.6\times 10^{3} \eta v_{8}^{4}M_{-4}^{-1}t_{\rm wk}^{3}\,\,{\rm s}
 &
{\rm FS}, \\
7.8\times 10^{4}  v_{8}^{4}\dot{M}_{-5}^{-1}t_{\rm wk}^{2} \,\,{\rm s} &
{\rm RS}, \\
\end{array}
\right.
\label{eq:tcool}
\end{eqnarray}
where $\Lambda = \Lambda_0 T^{1/2} = 2\times 10^{-27}T^{1/2}$ erg cm$^{3}$ s$^{-1}$ is the cooling function and we have assumed $\eta = 1$ in the RS case.  Our cooling function includes just free-free emission, which is conservative from the standpoint of radiative shocks because line cooling contributes a comparable or greater cooling rate at temperatures $\lesssim 3\times 10^{7}$ K (e.g.~\citealt{Schure+09}).  Here we have assumed that electrons and protons are efficiently coupled behind the shock, as is justified because the timescale for Coulomb energy exchange $t_{\rm e-p} = 14T_{\rm sh}^{3/2}/n_{\rm sh}$ s (NRL Plasma Formulary; \citealt{Huba07}) is short compared to the expansion timescale $t_{\rm exp} \sim R_{\rm ej}/v_{4} \sim t$,
\begin{eqnarray}
&& \frac{t_{\rm e-p}}{t}   \approx
\left\{
\begin{array}{lr}
  3\times 10^{-5}M_{-4}^{-1}v_{8}^{6}\eta^{3}t_{\rm wk}^{2} &
{\rm FS}, \\
1.2\times 10^{-3} \dot{M}_{-5}^{-1}v_{8}^{6}t_{\rm wk} &
{\rm RS}, \\
\end{array}
\right.
\label{eq:tep}
\end{eqnarray}  
where we have (conservatively) assumed a pure hydrogen composition.

Whether the shocks are radiative depends on the ratio of the cooling timescale to the expansion timescale,
\begin{eqnarray}
&&\chi \equiv \frac{t_{\rm cool}}{t} \approx
\left\{
\begin{array}{lr}
 2.7\times 10^{-3}\eta v_{8}^{4}M_{-4}^{-1}t_{\rm wk}^{2}
 &
{\rm FS}, \\
0.13  v_{8}^{4}\dot{M}_{-5}^{-1}t_{\rm wk}&
{\rm RS}, \\
\end{array}
\right.
\label{eq:cool}
\end{eqnarray}
For the characteristic range of parameters $M_{-4} \sim 0.3-3$ (\citealt{Seaquist&Bode08}), $v_{8} \lesssim 3$, $\eta < 1$ and timescales corresponding to the gamma-ray emission $t \lesssim $ few weeks, one concludes that the FS is likely to be radiative ($\chi < 1$).  It is less clear that the RS is radiative, since in principle for $v_{8} \gtrsim 2$ and a very low wind mass loss rate $\dot{M}_{-5} \lesssim 1$ one can have $\chi > 1$.  

The argument for radiative shocks can be strengthened by considering additional constraints.  First, the optical depth of the outer unshocked ejecta at optical frequencies, 
\be
\tau_{\rm opt} \simeq m_p n_{\rm ej}R_{\rm ej}\kappa_{\rm opt} \approx 0.9M_{-4}v_{8}^{-2}t_{\rm wk}^{-2},
\label{eq:tauopt}
\ee must exceed unity, where $\kappa_{\rm opt} \sim 0.1$ cm$^{2}$ g$^{-1}$ is the optical opacity set by Doppler-broadened iron lines (\citealt{Pinto&Eastman00}).  This constraint is motivated by the lack of clear evidence, e.g.~emission lines, for hot, energetic shocks in optically thin regions during the optical peaks of novae and during most epochs of gamma-ray detection.  Over the first few weeks or longer following outburst (i.e., prior to the drop of the ``iron curtain"), nova spectra are characterized by broad P-Cygni line profiles, indicating the presence of a pseudo-photosphere.  

Equation (\ref{eq:cool}) can be rewritten in terms of $\tau_{\rm opt}$ as
\begin{eqnarray}
&&\chi = \frac{9 m_p^{3/2}k^{1/2} \eta v_{\rm sh}^{2}\kappa_{\rm opt}}{8\sqrt{3}\Lambda_0 (n_{\rm sh}m_p R_{\rm ej}\kappa_{\rm opt})} \approx
\left\{
\begin{array}{lr}
 2\times 10^{-3}\eta v_{8}^{2}\tau_{\rm opt}^{-1}.
 &
{\rm FS}, \\
0.08 \dot{M}_{-5}^{-1}M_{-4}^{2}\tau_{\rm opt}^{-2}t_{\rm wk}^{-3}&
{\rm RS}. \\
\end{array}
\right.
\label{eq:tcooltau}
\end{eqnarray}
For the forward shock to occur below the photosphere ($\tau_{\rm opt} > 1$) yet not be radiative ($\chi > 1$) would thus require unphysically high ejecta velocities, $v_{4} \gtrsim 10,000$ km s$^{-1}$.  The RS is also radiative if $\tau_{\rm opt} > 1$ on timescales of a couple weeks for $M_{-4} \lesssim 3$ as long as $\dot{M}_{-5} \gtrsim 0.2$.    

A second requirement is that power dissipated by the shocks $L_{\rm sh} = 9\pi R_{\rm ej}^{2}n_{\rm sh}m_p v_{\rm sh}^{3}/32$ exceed the observed gamma-ray luminosity $L_{\gamma}$ by an efficiency factor $1/\epsilon_{\rm nth}\epsilon_{\gamma} \gtrsim 100$ accounting for the fraction of the shock power radiated as gamma-rays, where $\epsilon_{\rm nt}$ accounts for the efficiency of the shock in accelerating non-thermal particles and $\epsilon_{\gamma}$ accounts for the energy of the latter radiated in the LAT bandpass (see $\S\ref{sec:shockacceleration}$).  Equation (\ref{eq:cool}) can be written in terms of $L_{\rm sh} = L_{\gamma}/\epsilon_{\rm nth}\epsilon_{\gamma}$ as
\begin{eqnarray}
\chi &=&  \frac{81\pi}{256\sqrt{3}}\frac{\eta^{4}v_{4}^{5} m_p^{3/2} k^{1/2} R_{\rm ej} \epsilon_{\rm nth}\epsilon_{\gamma}}{L_{\gamma} \Lambda_0 } \nonumber \\
&\approx& 0.02\eta^{4} v_{8}^{6}t_{\rm wk}\left(\frac{L_{\gamma}}{10^{36}\,\rm erg\,s^{-1}}\right)^{-1}\left(\frac{\epsilon_{\rm nth}\epsilon_{\gamma}}{0.01}\right).
\label{eq:tcooltau}
\end{eqnarray}
For measured values $L_{\gamma} \sim 3\times 10^{35}$ erg s$^{-1}$ (\citealt{Ackermann+14}) the shocks are thus radiative ($\chi < 1$) on a timescale of weeks for $\epsilon_{\rm nth}\epsilon_{\gamma} < 0.01$ if the velocity of the shocks is $\lesssim 2,000$ km s$^{-1}$.

In conclusion, both forward and reverse shocks are likely to be radiative at times corresponding to the observed gamma-ray emission, although this statement is the most secure at the earliest times.  One can nevertheless keep the following arguments fully general by introducing the shock radiative efficiency $f_{\rm rad} =  (1 + 5\chi/2)^{-1}$, which equals unity for $\chi \ll 1$ but scales $\propto 1/\chi$ for $\chi \gg 1$ (\citealt{Metzger+14}).  


\subsection{The shock power will mostly emerge as optical radiation}
\label{sec:optical}

Absent the immediate presence of a shock, the bulk\footnote{Although the bulk of the ejecta is neutral at early times, X-rays from the shocks fully ionize a thin layer of gas just ahead of the shocks (\citealt{Metzger+14}).  Thus, we do not expect the shock to be modified by the presence of a neutral upstream medium, as may occur in some supernova remnants (e.g., \citealt{Blasi+12}).  } of the nova ejecta is neutral at early times because the timescale for radiative recombination, $t_{\rm rec} \sim 1/n_{\rm ej}\alpha_{\rm rec} \sim 11 v_{8}^{3}t_{\rm wk}^{3}M_{-4}^{-1}Z^{-2}$ s,  is extremely short compared to the evolution timescale $\sim$ weeks, where $\alpha_{\rm rec} \sim 10^{-12}Z^{2}$ cm$^{3}$ s$^{-1}$ is the approximate radiative recombination rate for hydrogen-like species of charge $Z$ (\citealt{Osterbrock&Ferland06}).  X-rays of temperature $T_{\rm sh} \lesssim 10^{7}$ K (eq.~[\ref{eq:Tsh}]) are thus absorbed by high columns of neutral gas ahead or behind the shock, before being re-radiated as line emission at lower frequencies.  If the shock is radiative then the timescale for reprocessing to lower temperatures is also necessarily short because the line cooling function $\Lambda(T)$ increases rapidly with decreasing temperature down to $T \sim 10^{4}$ K (e.g.~\citealt{Schure+09}).  However, harder X-rays with energies $\gtrsim 10$ keV should be free to escape even at these early times due the decreasing bound-free cross section at high photon frequencies ($\S\ref{sec:discussion}$).

Radiation escapes efficiently once it reaches the optical/near-UV band due to the much lower opacity at these frequencies compared to the UV/X-ray opacity of neutral gas.  Reprocessed optical radiation will furthermore not be degraded by PdV losses provided that the photon diffusion timescale $t_{\rm d} \sim R_{\rm ej}\tau_{\rm opt}/c$ is shorter than the expansion timescale $t_{\rm exp} \sim R_{\rm ej}/v_{4} \approx t$, where $\tau_{\rm opt}$ is the optical depth (eq.~[\ref{eq:tauopt}]).  Equating these two, optical radiation escapes without adiabatic losses after a time (\citealt{Arnett82})
\be
t_{\rm opt} \approx 0.4 M_{-4}^{1/2}v_{8}^{-1/2}{\rm d}.
\label{eq:topt}
\ee  
The timescale $t_{\rm opt}$ also sets the minimum timescale for the optical lightcurve to rise after the onset of the explosion.

The onset of gamma-ray emission from V1324 Sco and V339 Del was delayed with respect to the peak of the optical radiation by a few days (\citealt{Ackermann+14}).  The dominant process\footnote{Photo-pion production is the dominant source of opacity for higher energy gamma-rays $\epsilon_{\gamma} \gtrsim $ 3 GeV, for which the opacity is $\kappa_{\pi} \sim 3\times 10^{-4} \kappa_{\rm es}$  (e.g.~\citealt{Anchordoqui+02}, \citealt{Montanet+94}).  Photon-photon pair creation opacity becomes important at TeV energies (eq.~[\ref{eq:TeV}]).} by which gamma-rays of energy $\epsilon_{\gamma} \sim 0.1-3$ GeV are attenuated is inelastic electron scattering, for which the Klein-Nishina opacity is $\kappa_{\rm ie} \sim 10^{-3}(\epsilon_{\gamma}/{\rm GeV})^{-1}\kappa_{\rm es} \sim 4\times 10^{-3}(\epsilon_{\gamma}/{\rm GeV})^{-1}\kappa_{\rm opt}$.  The ejecta will thus remain opaque to gamma-rays of energy $\epsilon_{\gamma}$ until $\tau_{\rm opt} \lesssim 250(\epsilon_{\gamma}/{\rm GeV})$, as occurs after a time $t_{\gamma}$ which exceeds that of the nominal optical rise time (eq.~[\ref{eq:topt}]) by the ratio
\be
\frac{t_{\gamma}}{t_{\rm opt}} \approx \left(\frac{c \kappa_{\rm ie}}{v_{\rm w}\kappa_{\rm opt}}\right)^{1/2} \approx 1.1\left(\frac{\epsilon_{\gamma}}{{\rm GeV}}\right)^{-1/2}v_{8}^{-1/2},
\ee
i.e.~$t_{\gamma} \sim $ couple days for $\epsilon_{\gamma} \sim $0.1 GeV.  

From the above we can draw two key conclusions: (i) a shock that produces gamma-ray emission which is not absorbed and hence observable (gamma-ray optical depth $\tau_{\gamma} < 1$) necessarily radiates the bulk of its dissipated thermal energy without adiabatic losses at optical frequencies ($\tau_{\rm opt} \lesssim c/v_{4}$).  In other words, at times when the column of ejecta is sufficiently low to allow gamma-rays to escape unattenuated, it is also sufficiently low to allow thermal radiation to escape without significant losses to PdV work. (ii) gamma-ray absorption ($\tau_{\gamma} > 1$) at early times may explain the delayed onset of the gamma-ray emission.  If true, the latter implies that the optical emission near peak can be shock-powered, even if gamma-ray emission is suppressed at this time.  Alternatively, the gamma-ray delay may result from the finite timescale required to accelerate the gamma-ray producing particles at the shock (G.~Dubus, private communication).

\subsection{Partitioning the shock energy}
\label{sec:shockacceleration}

A large fraction $f_{\rm rad} \sim 1$ of the total power $L_{\rm sh}$ dissipated by shocks goes into thermal X-rays, which are absorbed and re-radiated as optical emission ($\S\ref{sec:optical}$).  A much smaller fraction $\epsilon_{\rm nth} \ll 1$ goes into accelerating non-thermal ions or electrons.  The fraction of non-thermal power $\epsilon_{\rm nth}$ radiated as gamma-rays, $L_{\gamma} = \epsilon_{\rm nth}\epsilon_{\gamma}L_{\rm sh}$, depends also on a factor $\epsilon_{\gamma} <1 $ accounting for the radiative efficiency of the accelerated particles and the fraction of the total gamma-ray emission emitted in the LAT bandpass.  Combining these expressions, the fraction of the total nova optical luminosity powered by shocks can be written
\be
f_{\rm sh} \equiv \frac{L_{\rm opt,sh}}{L_{\rm opt}} \approx \frac{f_{\rm rad}}{\epsilon_{\gamma}\epsilon_{\rm nth}}\frac{L_{\gamma}}{L_{\rm opt}}.
\label{eq:fsh}
\ee
Once $\epsilon_{\gamma}$ is specified based on the assumed emission process (hadronic or leptonic), the observed ratio $L_{\gamma}/L_{\rm opt}$ sets a lower limit on the value of $\epsilon_{\rm nth}$, i.e. 
\be
f_{\rm sh} < 1 \Rightarrow \epsilon_{\rm nth} > \epsilon_{\rm nth,min} = \frac{1}{\epsilon_{\gamma}}\frac{L_{\gamma}}{L_{\rm opt}},
\label{eq:entmin}
\ee
assuming a radiative shock ($f_{\rm rad} = 1$).  Alternatively, if the value of $\epsilon_{\rm nth}$ is assumed, then the measured value of $L_{\gamma}/L_{\rm opt}$ determines $f_{\rm sh}$.  We now consider what values of $\epsilon_{\rm nth}$ and $\epsilon_{\gamma}$ are expected in hadronic and leptonic scenarios, respectively.

\subsubsection{Hadronic scenario}
\label{sec:hadronic}

The momentum distribution $f(p) \propto p^{-4}$ of non-thermal protons accelerated via Diffusive Shock Acceleration (e.g., \citealt{Blandford&Ostriker78}) corresponds to an energy distribution
\begin{eqnarray}
&& \frac{dN_p}{dE_p}E_p^{2} \propto  
\left\{
\begin{array}{lr}
E_p^{1/2}
, &
kT_{\rm sh} \lesssim E_p \ll m_p c^{2} \\
{\rm constant,} &
m_p c^{2} \ll E_p < E_{\rm max}, \\
\end{array}
\right..
\label{eq:dNdEion}
\end{eqnarray}
that concentrates most of the non-thermal energy in relativistic particles.  The \citet{Hillas84} criterion sets an upper limit on the maximum proton energy
\be
E_{\rm max} = \frac{e B_{\rm sh} v_{\rm sh}R_{\rm ej}}{c} \approx 7\times 10^{12}\,{\rm eV}\,\eta^{2} \left(\frac{\epsilon_{B}}{10^{-6}}\right)^{1/2}M_{-4}^{1/2}v_{8}^{3/2}t_{\rm wk}^{-1/2},
\label{eq:Emax}
\ee
where $B_{\rm sh}$ is the magnetic field strength near the shock, which is estimated by assuming that magnetic pressure $B_{\rm sh}^{2}/8\pi$ is a fraction $\epsilon_{\rm B}$ of the thermal pressure of the post-shock gas.  

A reasonably high value of $\epsilon_{B}$ is expected if ion acceleration is efficient, because cosmic-ray induced instabilities amplify the magnetic field for some distance behind the shock to a significant fraction of its equipartition value.  \citet{Caprioli&Spitkovsky14b} show that the magnetic field near the shock is amplified from its initial upstream value $B_0$, according to $(B_{\rm sh}/B_{0})^{2} \approx 3\epsilon_{\rm nth}M_{A}$ (their eq. 2), where $M_{A} = v_{\rm sh}/v_{A}$ is the Alfv\'enic Mach number of the upstream\footnote{For concreteness, we focus here on amplification of the magnetic field of the slow ejecta at the forward shock, although similar considerations apply to amplification of the field in the fast outflow at the reverse shock.} flow in the shock frame, $v_{A} = B_{0}/\sqrt{4\pi m_p n_{4}}$, and $\epsilon_{\rm nth}$ is the fraction of the shock dissipated kinetic energy placed into relativistic ions.  The equipartition fraction of the shock region can thus be written as \be
\epsilon_{B} =   3\epsilon_{\rm nth}\eta^{-1}\epsilon_{B,0}^{1/2} \approx 10^{-6}\eta^{-1}\left(\frac{\epsilon_{\rm nth}}{0.1}\right)\left(\frac{\epsilon_{B,0}}{10^{-11}}\right)^{1/2},
\ee
where $\epsilon_{B,0} \equiv B_{0}^{2}/(4\pi n_4 m_p v_{\rm 4}^{2})$ is the ratio of magnetic energy to kinetic energy within the initial, unshocked outflow.

The value of $\epsilon_{B,0}$ within the unshocked nova ejecta can be constrained as follows.  Assume that the magnetic energy density of the outflow $U_{\rm B}$ is a fraction $\epsilon_{B,\star}$ of the bulk kinetic energy $U_{\rm K} \propto \rho v^{2}/2$ near the stellar surface ($r = R_{\star}$), where $\rho$ and $v$ are the outflow  density and velocity, respectively.  Flux-freezing within the 3D expanding flow ($\rho \propto r^{-3}$) dilutes the field strength according to $B \propto 1/r^{2}$, such that $U_{\rm B}/U_{\rm k} = \epsilon_{\rm B,0}$ at the radius of the shock $\sim R_{\rm ej}$ will be at most a factor of $R_{\star}/R_{\rm ej}$ smaller than $\epsilon_{B,\star}$.  For characteristic values of $R_{\rm ej} \sim 10^{14}$ cm and the initial radius $R_{\star} \gtrsim 10^{9}$ cm of the outflow set by the white dwarf surface, we expect that $\epsilon_{B,0} \gtrsim 10^{-5}\epsilon_{B,\star}$.  Thus, for $\epsilon_{\rm nth} = 0.1$ and $\eta = 1$, even an extremely small value of $\epsilon_{B,\star} \gtrsim 10^{-18}$ appears sufficient to accelerate protons to a maximum energy $E_{\rm max} \gtrsim 10^{10}$ eV (eq.~[\ref{eq:Emax}]) large enough to explain the highest LAT-detected photon energies via $\pi_0$ creation (\citealt{Ackermann+14}).  

We caution, however, that the bulk of the ejecta is neutral and hence may not be able to support the strong turbulent magnetic field needed to accelerate particles.  In this case the radius $R_{\rm ej}$ entering the Hillas criterion in equation (\ref{eq:Emax}) should be replaced by the width of the X-ray ionized layer ahead of the shock. \citet{Metzger+14} estimate the latter to be $\Delta^{\rm ion} \sim 9\times 10^{9}\eta v_{8}(n_{\rm ej}/10^{10}{\rm \,cm^{-3}})^{-1}$ cm based on ionization by the free-free emission (their eq.~[45]).  Replacing $R_{\rm ej}$ with $\Delta_{\rm ion}$ in equation (\ref{eq:Emax}) results in 
\begin{eqnarray}
E_{\rm max} &=&  \frac{e B_{\rm sh} v_{\rm sh}\Delta_{\rm ion}}{c} \nonumber \\
&\approx& 1.0\times 10^{9}\,{\rm eV}\,\eta^{3} \left(\frac{\epsilon_{B}}{10^{-6}}\right)^{1/2}\left(\frac{n_{\rm ej}}{10^{10}\,{\rm cm^{-3}}}\right)M_{-4}^{1/2}v_{8}^{3/2}t_{\rm wk}^{-3/2}.
\label{eq:Emax2}
\end{eqnarray}
Thus, for typical values of $n_{\rm ej} \sim 10^{9}-10^{10}$ cm$^{-3}$ (eq.~[\ref{eq:nej}], [\ref{eq:nw}]) and $v_{8} \sim 1$, one finds that higher field strengths of $\epsilon_{B} \gtrsim 10^{-2}-10^{-4}$ ($\epsilon_{B,\star} \gtrsim 10^{-2}-1$) is now required to have $E_{\rm max} \gtrsim 10^{10}$ eV.  Note, however, that \citet{Metzger+14} do not take into account additional ionization of hydrogen by the UV radiation produced by line cooling or by the reprocessed X-rays, which could be important at early times of relevance because the ejecta is opaque to X-rays.  This could increase the energy of the ionizing radiation, and hence the thickness of the ionized layer and $E_{\rm max}$, by up to a factor of $\sim kT_{\rm sh}$/(2 Ryd) $\sim 50 \eta^{2}v_{8}^{2}$, where Ryd = 13.6 eV is the characteristic ionization threshold energy for hydrogen, relaxing constraints on $\epsilon_{B,\star}$.

The fraction of the shock power placed into non-thermal ions $\epsilon_{\rm nth}$ can be estimated from observations of other non-relativistic shocks, such as gamma-ray emission in supernova remnants (e.g.~\citealt{Ackermann+14}).  In Tycho, for instance, \citet{Morlino&Caprioli12} infer $\epsilon_{\rm nth} \sim 0.1$ if the gamma-rays are hadronic in origin.  Hybrid (kinetic ions - fluid electron) simulations of non-relativistic shocks also find $\epsilon_{\rm nth} \approx 0.1-0.2$ for cases in which the upstream magnetic field is quasi-parallel to the shock normal (\citealt{Caprioli&Spitkovsky14}).  However, if the nova ejecta is characterized by a phase of approximately steady-state outflow, then radial expansion in two dimensions will produce a magnetic field dominated by its toroidal component, i.e. perpendicular to the outflow and hence shock direction (Fig.~\ref{fig:schematic}).  For such quasi-perpendicular magnetic field geometries, \citet{Caprioli&Spitkovsky14} infer much lower proton acceleration efficiencies (consistent with zero), a point we return to in $\S\ref{sec:discussion}$. 

Relativistic protons accelerated at the shocks produce pions by colliding with effectively stationary protons in the ejecta.  The mean time between interactions is given by $t_{\rm p-p} = (\kappa_{\rm p-p}m_p n_{\rm ej} c)^{-1}$, where $\kappa_{\rm p-p} \sim 0.025$ cm$^{-2}$ g$^{-1} \approx \kappa_{\rm opt}/4$ (\citealt{Kamae+06}) is the opacity for inelastic proton collisions of energy $E_{\rm p} \gtrsim $ GeV.  The number of collisions a proton experiences over an expansion timescale $t/t_{\rm p-p}$ thus exceeds unity until after a time
\be
t_{\rm p-p} \approx 8.2\,M_{-4}^{1/2}v_{8}^{-3/2}\,{\rm weeks}.
\label{eq:tpp}
\ee
At times $t \ll t_{\rm p-p}$, protons lose their energy to pion production instead of adiabatic expansion; the ejecta thus acts as an efficient ``calorimeter" for converting relativistic protons into gamma-rays.

The fraction of $p-p$ collisions producing gamma-rays $\epsilon_{\gamma}$ is the product of the fraction of inelastic interactions, $\sigma_{\rm ie}/(\sigma_{\rm ie} + \sigma_{\rm e}) \sim 0.5-0.8$ (where $\sigma_{\rm e}$ and $\sigma_{\rm ie}$ are the elastic and inelastic cross sections, respectively; \citealt{Kamae+06}), and the fraction $\simeq 1/3$ of inelastic events placed into the $p + p \rightarrow \pi^{0} \rightarrow \gamma + \gamma$ channel.  We thus typically expect that $\epsilon_{\gamma} \approx 0.2-0.3$, depending on the details of the accelerated proton spectrum, while a lower value could result if protons are not efficiently trapped, i.e. at times $t > t_{\rm p-p}$.  

To summarize, we expect $\epsilon_{\rm nth}\epsilon_{\gamma} \lesssim 0.03$ in hadronic scenarios if the magnetic field within the ejecta is perpendicular to the shock plane, but this value may be considerably lower if the field is instead parallel to the shock plane, as would be expected given a phase of quasi-steady outflow from the white dwarf surface.



\subsubsection{Leptonic scenario}
\label{sec:leptonic}

In supernova shocks, the ratio of the energy placed into relativistic protons to that in relativistic electrons is estimated to be $K_{\rm ep} \sim 10^{-4}-10^{-2}$ based on observations of individual remnants (\citealt{Volk+05}; \citealt{Morlino&Caprioli12}) and synchrotron emission from Galactic cosmic rays (\citealt{Beck&Krause05}; \citealt{Strong+10}).  This value is consistent with recent particle-in-cell simulations of non-relativistic shocks, which find $K_{\rm ep} \sim 10^{-3}$ when extrapolated to shock velocities $v_{\rm sh}/c \lesssim 0.01$ characteristic of those in novae (\citealt{Park+14}; \citealt{Kato14}).  Assuming shocks accelerate protons with a maximum efficiency $\sim 0.1$ (\S \ref{sec:hadronic}), $K_{\rm ep} \lesssim 10^{-2}$ corresponds to a relativistic non-thermal electron fraction of $\epsilon_{\rm nth} \lesssim 10^{-3}$.

To Compton upscatter optical seed photons of energy $E_{\rm opt} \sim $ eV energy to $\sim 0.1-10$ GeV requires electrons with energy $E_e = \gamma_e m_e c^{2}$ and Lorentz factor $\gamma_e \sim 10^{4}-10^{5}$.  The ratio of the Compton cooling timescale of the electron to the expansion timescale is given by
\be
\frac{t_{\rm cool,IC}}{t} = \frac{3 m_e c  }{4\sigma_T  U_{\rm opt} \gamma_e t} \sim 0.07 \tau_{\rm opt}^{-1}\left(\frac{\gamma_e}{10^{4}}\right)^{-1}\left(\frac{L_{\rm opt}}{10^{38}\,\rm erg\,s^{-1}}\right)^{-1}v_{8}^{2}t_{\rm wk},
\label{eq:tcool2}
\ee
where $U_{\rm opt} \approx F\tau = L_{\rm opt}\tau_{\rm opt}/4\pi c R_{\rm ej}^{2}$ is the radiation energy density near the shock (for shocks below the photosphere, $\tau_{\rm opt} > 1$), where we have used the fact that the conserved radiative flux $F \propto \partial U_{\rm opt}/\partial \tau$ in the diffusion approximation implies that $F \tau$ is constant from the photosphere ($\tau \sim 1$) to the shock depth $\tau = \tau_{\rm opt}$.  Equation (\ref{eq:tcool2}) shows that $\gamma_e \gtrsim 10^{4}$ electrons are in the fast cooling regime ($t_{\rm cool} \ll t_{\rm exp}$) for luminosities $L_{\rm opt} \gtrsim 10^{37}$ erg s$^{-1}$ and timescales $t \sim$ weeks of relevance.  

Relativistic Bremsstrahlung emission represents an alternative, and possibly dominant, leptonic emission mechanism.  The cooling rate of a relativistic electron of Lorentz factor $\gamma_e$ off protons in the nova ejecta is given by $\dot{q}_{\rm rb} = 3\times 10^{-22}\gamma_e^{3/2}n_{\rm ej}$ erg s$^{-1}$ (\citealt{Rybicki&Lightman79}), resulting in a cooling time
\be
\frac{t_{\rm cool,rb}}{t} = \frac{\gamma_e m_e c^{2}}{\dot{q}_{\rm rb}} \approx 5.1\times 10^{-4} \left(\frac{\gamma_e}{10^{4}}\right)^{-1/2}M_{-4}^{-1}t_{\rm wk}^{2} v_{8}^{3} ,
\label{eq:tcool2B}
\ee
where we have used equation (\ref{eq:nej}).  A comparison between equations (\ref{eq:tcool2}) and (\ref{eq:tcool2}) shows that relativistic bremsstrahlung emission can exceed that from Inverse Compton scattering for typical values of the nova luminosity, outflow velocity, and ejecta mass.  Regardless of whether Inverse Compton or bremsstrahlung emission dominates, the shock energy placed into electrons with the necessary Lorentz factors to produce the observed gamma-ray emission will in fact be radiated with high efficiency.

Compton scattering of electrons with an accelerated spectrum $dN_e/dE_e \propto E_e^{-p}$, where $E_e = \gamma_e m_e c^{2}$, produces a gamma-ray spectrum 
\begin{eqnarray}
&& \frac{dN_\gamma}{dE} \propto  
\left\{
\begin{array}{lr}
E^{-(p+1)/2}
, &
 E \ll  E_{\rm c}  \\
E^{-(p+2)/2} &
E \gg E_{\rm c}, \\
\end{array}
\right..
\label{eq:dEgammae}
\end{eqnarray}
where $E_{\rm c} \sim E_{\rm opt}\gamma_{\rm c}^{2}$ and $\gamma_{\rm c}$ is the minimum electron Lorentz factor obeying the fast-cooling condition ($t_{\rm cool,IC} \lesssim t_{\rm exp}$; eq.~[\ref{eq:tcool2}]).  A similar cooling energy can be defined in the case of relativistic brehmsstrahlung emission, below which the photon spectrum $dN_{\gamma}/dE \propto E^{-(2p-1)/2}$, i.e. the same as for Inverse Compton in the $p = 2$ case of greatest relevance (see below).

The fraction of the total energy in relativistic electrons radiated in the LAT bandpass is given by
\begin{eqnarray}
&& \epsilon_{\gamma} < \frac{\int_{10^{4}}^{10^{5}}\frac{dN_{e}}{dE_e}E_e dE_e }{\int_{1}^{10^{5}}\frac{dN_{e}}{dE_e}E_e dE_e } \approx
\left\{
\begin{array}{lr}
10^{8-4p}-10^{10-5p} \ll 0.2
, &
 p > 2 \\
0.2 &
p = 2, \\
\end{array}
\right..
\label{eq:egammae}
\end{eqnarray}
where we have assumed that  that particles with $10^{4} \lesssim \gamma_e \lesssim 10^{5}$ are the only ones contributing into the LAT bandpass and that they radiative with 100 percent efficiency, i.e. $t_{\rm cool,IC} \ll t$ or $t_{\rm cool, rb} \ll t$.  Power-law fits to the gamma-ray spectra of three classical novae yield best-fit photon indices $\Gamma \sim 2-2.5$ (\citealt{Ackermann+14}) corresponding to $p = 2(\Gamma -1) \sim 2-3$, for which we estimate $\epsilon_{\gamma} \sim 0.2-10^{-4}$.  To be conservative we hereafter assume $p = 2$, corresponding to $\epsilon_{\gamma} = 0.2$.

In summary, adopting $\epsilon_{\rm nth} \sim 10^{-5}-10^{-3}$ motivated by simulations and observations, we estimate that $\epsilon_{\rm nth}\epsilon_{\gamma} \sim 10^{-6}-10^{-4}$ in the leptonic scenario. 

\section{Data}
\label{sec:data}

This section describes our analysis of the gamma-ray and optical light curves of V1324 Sco and V3229 Del, events chosen as currently being the only classical novae with both published gamma-ray data and contemporaneous optical coverage.  We do not consider symbiotic novae in this analysis because the argument for the shocks being radiative is less secure given that the latter propagate into the extended wind of the giant companion star, instead of the potentially denser slow ejecta in the case of internal shocks.  Our goal is to use the measured ratio of gamma-ray and optical fluxes to constrain the particle acceleration efficiency $\epsilon_{\rm nth}$ and the fraction of the classical nova optical light curve powered by shocks  $f_{\rm sh}$, following the arguments outlined in $\S\ref{sec:shockacceleration}$.   

\subsection{Optical}

For each nova we use photometric measurements at both optical and near-infrared (NIR) wavelengths.  Optical photometry was taken from the database of the American Association of Variable Star Observers \citep{website:aavso}. Both nova Sco and Del had $BVR$ band measurements, but only Del included $I$ band. To supplement the NIR for nova Sco, we used $JHK$ measurements from the Small and Moderate Aperture Research Telescope (F.~M.~Walter, private communication).  Only photometric measurements coincident with the published Fermi gamma-ray light curve were used; specifically August 16$-$September 14 2013 and June 15$-$June 30 2012 for V339 Del and V1324 Sco, respectively. Photometric reddening corrections were applied to the datasets using an $E(B-V) = 0.2$ for V339 Del~\citep{Munari:2013p3504} and $E(B-V) = 1.0$ for V1324 Sco~\citep{Finzell:2015}. Filter specific reddening corrections for the optical/NIR were taken from \cite{Schlafly:2011p3366} assuming $R_V = 3.1$.

\begin{figure}
\subfigure{
\includegraphics[width=0.5\textwidth]{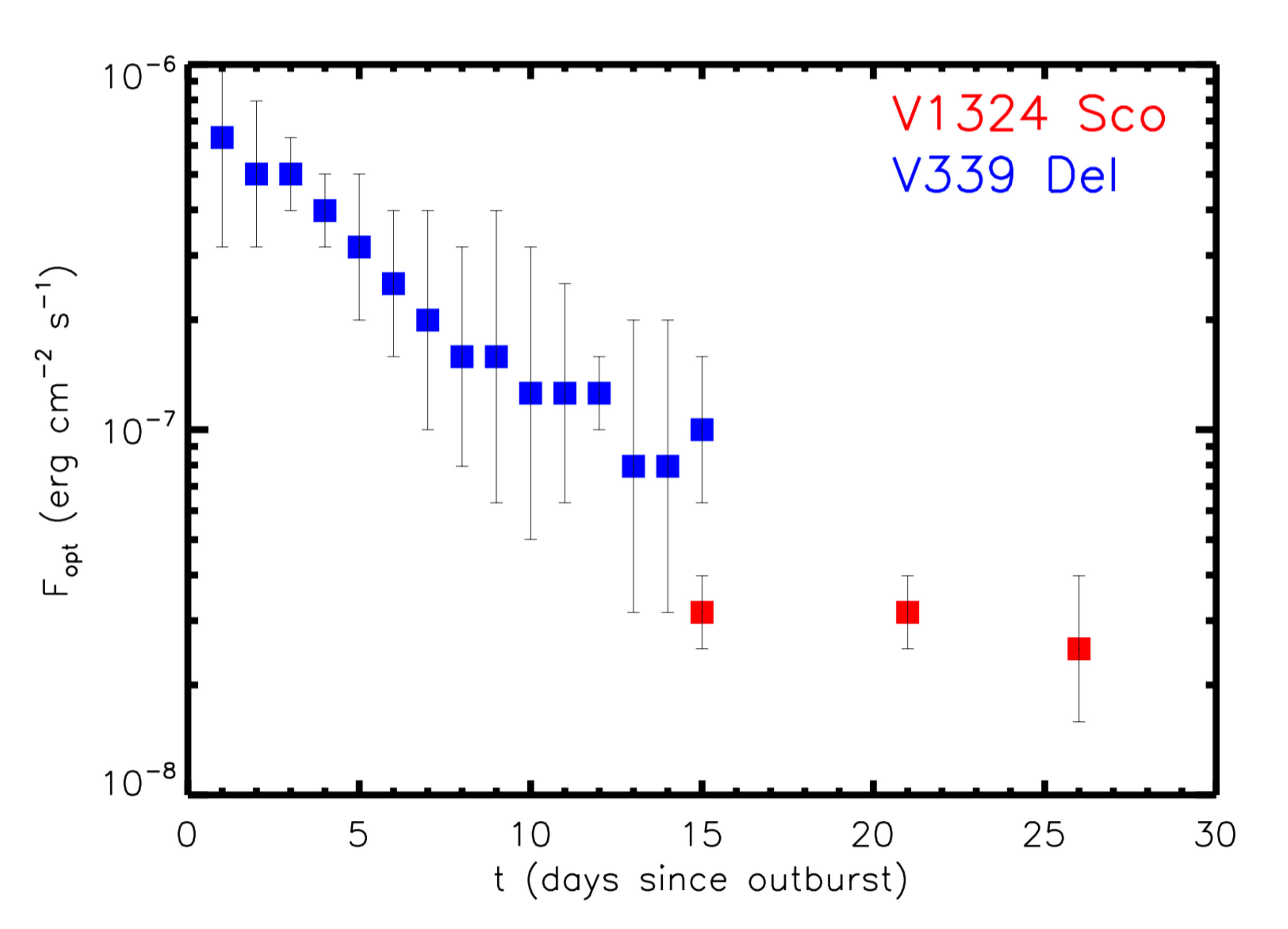}}
\subfigure{
\includegraphics[width=0.5\textwidth]{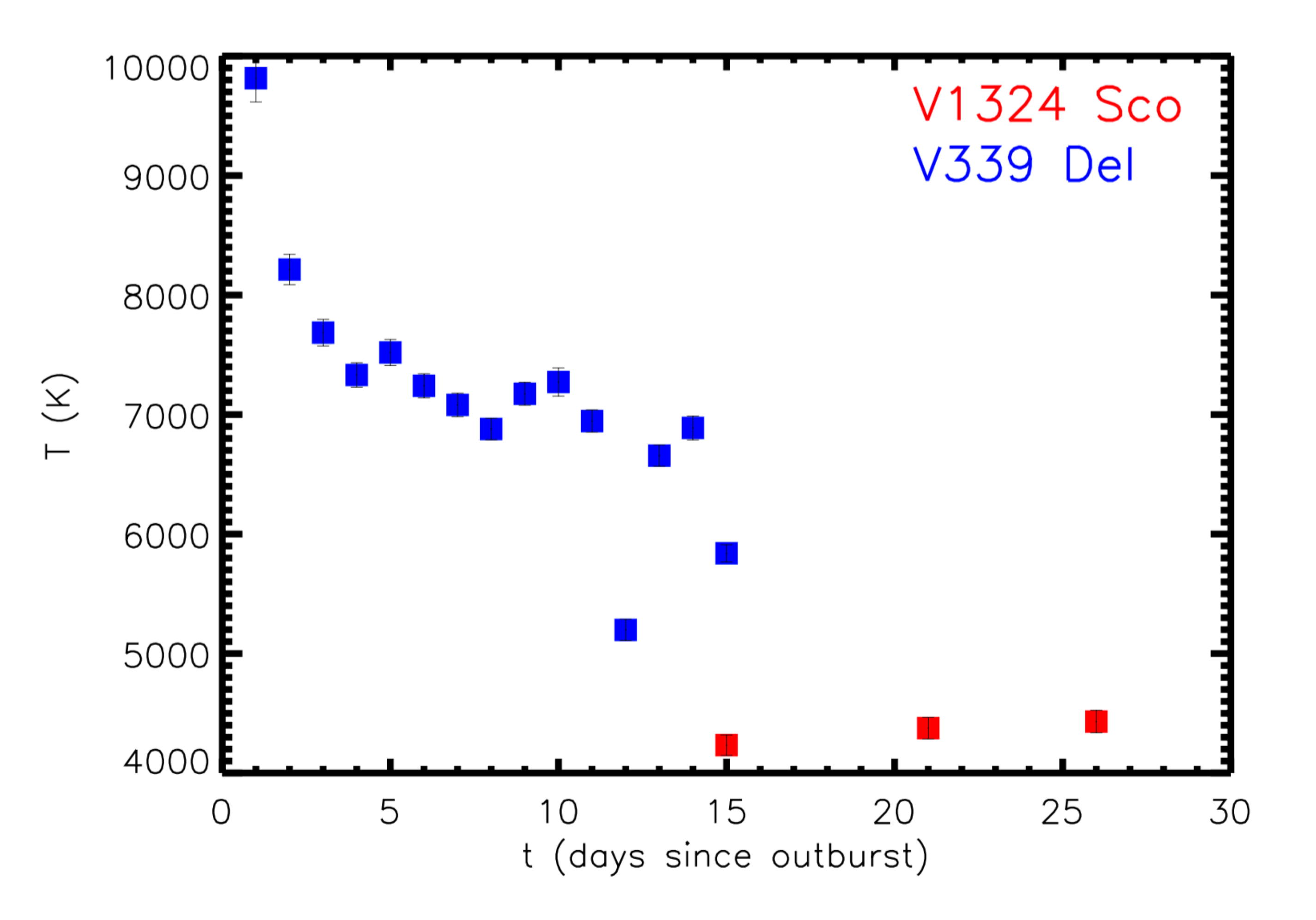}}
\caption{Total optical flux ({\it top panel}) and best-fit blackbody temperature ({\it bottom panel}) as a function of time since outburst for novae V1324 Sco ({\it blue}) and V339 Del ({\it red}).  Time is measured starting on June 1, 2012 for Nova Sco (beginning of optical outburst), and starting on August 16, 2013 for Nova Del (epoch of first gamma-ray detection, within days of the optical rise) .  }
\label{fig:fluxtemp}
\end{figure}

The total optical/NIR flux each night, $F_{\rm opt}$, was determined by approximating the SED as a blackbody, fitting a temperature and normalization, and integrating over frequency. The early-time flux obtained for V339 Del are consistent with the values found in the more detailed analysis of~\citet{Skopal:2014p3414}, who also showed that the spectral shape was a blackbody all the way out to the mid-IR.  We confirmed our blackbody assumption for V1324 Sco by extending our spectral fit up to the near-UV and comparing with flux values from \emph{Swift} UVOT~\citep{Page:2012p3257}. After applying reddening corrections from \citet{Brown:2010p3537} our blackbody fit predicts a near-UV flux of $2.1\pm 0.4 \times 10^{-13}$ erg s$^{-1}$ cm$^{-2}$, a value in agreement with the observed flux, $2.4\pm 0.16 \times 10^{-13}$ erg s$^{-1}$ cm$^{-2}$. As the near-UV flux is a factor $\sim 10^5$ times lower than in the optical, this excellent agreement validates the assumed blackbody spectral profile.  Results for the total optical/NIR flux and best-fit temperature as a function of time are shown in Figure \ref{fig:fluxtemp}.

As only three nights have complete NIR + BVR color data in nova Sco, we also consider separately the flux ratio on other nights with LAT detections obtained using just the V magnitude to calculate the optical/NIR flux, assuming a bolometric correction identical to that measured with better color data on day 15.  Uncertainty in the total flux is estimated by combining the uncertainty in the best-fit constant offset with an uncertainty in the temperature fit derived using the error in individual flux measurement.

\subsection{Gamma-Ray}
The {\it Fermi} LAT data for nova Sco and Del are taken from \citet{Ackermann+14}.  Spectral fits to the gamma-ray emission (over its entire duration) are provided by these authors for two profiles: Power Law (PL, $dN_{\gamma}/dE \propto E^{-\Gamma}$) and Power Law with Exponential Cut-Off at energy $E = E_{\rm c}$ (EPL, $dN_{\gamma}/dE \propto E^{-s} \exp^{-E/E_c}$).  The best-fit values of the free parameters, $\Gamma$ (PL) or $s$ and $E_{\rm c}$ (EPL), are provided along with their uncertainties.  

Also provided for each day is the average photon number flux above $100$ MeV, $N_{\gamma} \propto \int_{100\,\mathrm{ MeV}}^{100\,\mathrm{ GeV}} \frac{dN_{\gamma}}{dE} dE $.  The energy flux $F_{\gamma} \propto \int_{100\mathrm{ MeV}}^{100\,\mathrm{ GeV}} E\frac{dN_{\gamma}}{dE} dE$ is calculated from $F_{\gamma}$ according to
\begin{equation}
F_{\gamma} = \frac{N_{\gamma} \int_{100\,\mathrm{ MeV}}^{100\,\mathrm{ GeV}}E  \frac{dN_{\gamma}}{dE} dE}{\int_{100\,\mathrm{ MeV}}^{100\,\mathrm{  GeV}} \frac{dN_{\gamma}}{dE} dE},
\end{equation}
where uncertainties in $F_{\gamma}$ are derived from both the quoted uncertainties in $N_{\gamma}$ and in the best-fit parameters of the spectral fits.  Although we calculate the gamma-ray flux each day, our analysis assumes a constant spectral shape over the duration of the outburst, as necessitated by the poor statistics of the gamma-ray detections.    

\subsection{Flux Ratio}

\begin{figure}
\subfigure{
\includegraphics[width=0.5\textwidth]{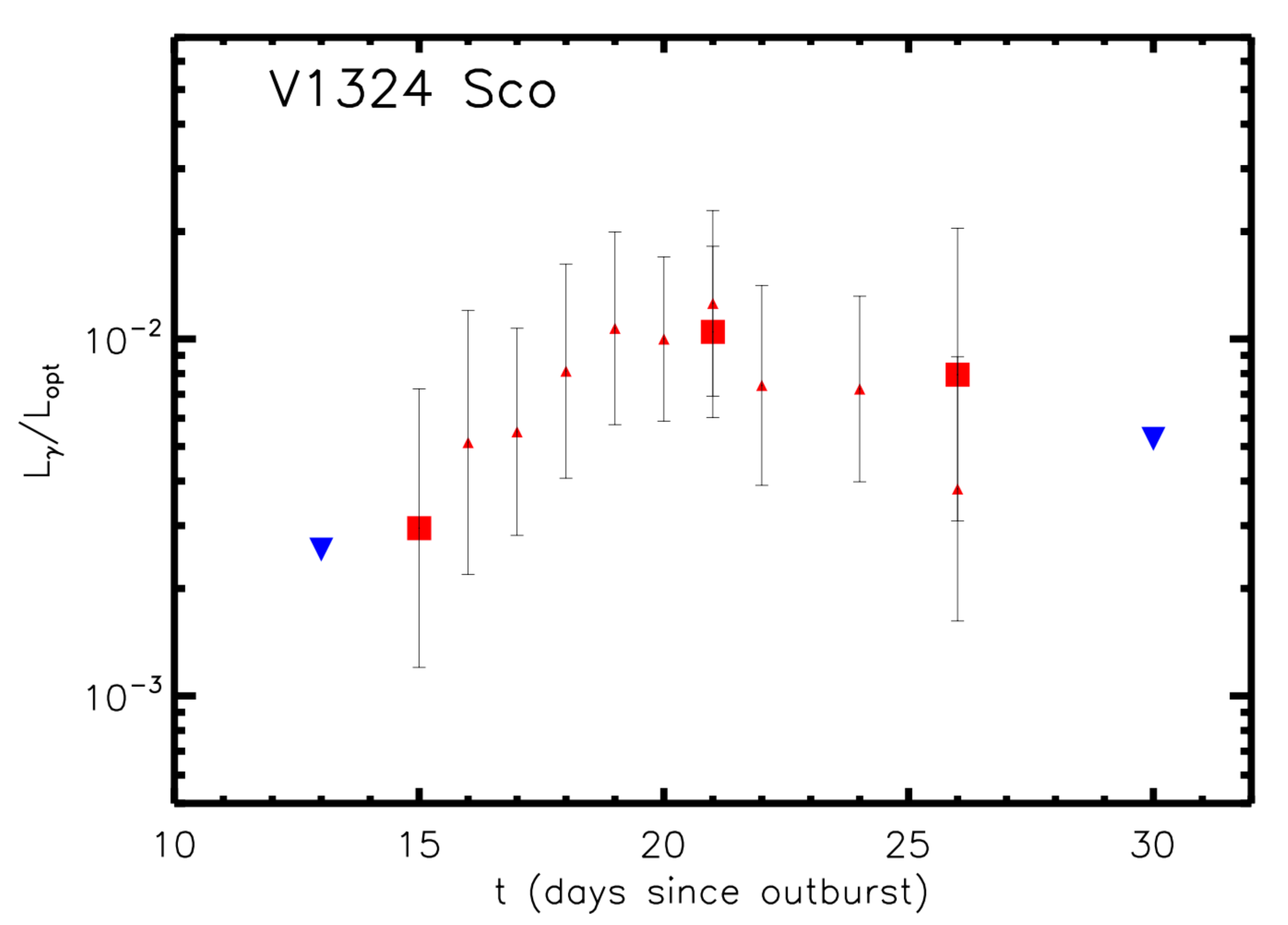}}
\subfigure{
\includegraphics[width=0.5\textwidth]{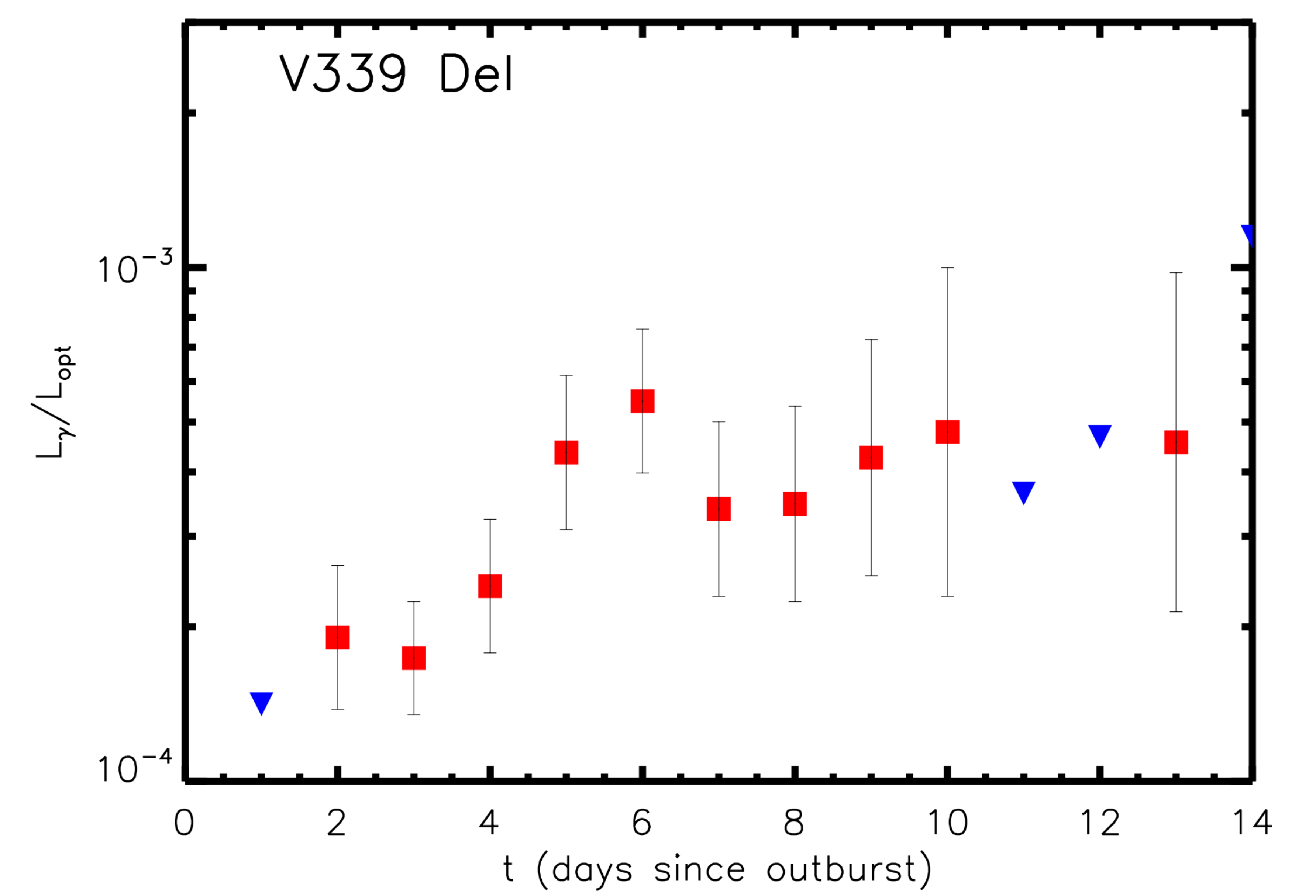}}
\caption{Ratio of gamma-ray and optical luminosities, $L_{\gamma}/L_{\rm opt}$, for novae V1324 Sco ({\it top panel}) and V339 Del ({\it bottom panel}) as a function of time since outburst.  In Nova Sco, large squares show nights with full NIR+VBR coverage, while small triangles show the flux ratio estimated using just the V-band magnitude adopting the same bolometric correction as measured with the full data color data on Day 15.  }
\label{fig:ratio}
\end{figure}

Our results for the flux ratio $L_{\gamma}/L_{\rm opt} = F_{\gamma}/F_{\rm opt}$ as a function of time for nova Sco and Del are shown in Figure \ref{fig:ratio}.  Key results are summarized in Table \ref{table:data}.  The ratio $L_{\gamma}/L_{\rm opt}$ is consistent with being constant in time in nova Sco, but nova Del shows evidence for a moderate secular increase.

We find an average value $L_{\gamma}/L_{\rm opt} \approx 10^{-2.2}(10^{-3.5})$ for nova Sco(Del), respectively.  Assuming radiative shocks and gamma-ray efficiency $\epsilon_{\gamma} \lesssim 0.2$ ($\S\ref{sec:hadronic}, \ref{sec:leptonic}$), the minimum non-thermal particle acceleration efficiency corresponds to $\epsilon_{\rm nth} \gtrsim  0.1-0.01$ in nova Sco and $\epsilon_{\rm nth} \gtrsim  10^{-3}$ in nova Del.  This greatly exceeds the acceleration efficiency expected for electron acceleration at non-relativistic shocks, disfavoring lepton scenarios for the gamma-ray emission.  Furthermore, assuming a proton acceleration efficiency of $\epsilon_{\rm nth} \sim 0.01-0.1$, we conclude that a fraction $f_{\rm sh} \gtrsim 10-100$ per cent of the optical emission was shock-powered in nova Sco.  The shock-powered fraction is $f_{\rm sh} \gtrsim 1-10$ per cent in nova Del. 

\begin{table}
\centering
\begin{minipage}{16cm}
\caption{$\gamma$-ray/optical flux ratio in classical novae and implications.}
\begin{tabular}{lccc}
\hline
{Nova}&
{$\langle$ log$\left[\frac{L_{\gamma}}{L_{\rm opt}}\right] \rangle$} & 
{log$\epsilon_{\rm nth,min}^{(a)}$} &
{$f_{\rm sh}^{(b)}$} \\
\hline
V1324 Sco & -2.2$\pm 0.4$  & -1.5$\pm0.4$ & 0.13($\epsilon_{\rm nth}/0.1)^{-1}$ \\
V339 Del  & -3.5$\pm 0.2$  & -2.8$\pm 0.2$ & 0.016($\epsilon_{\rm nth}/0.1)^{-1}$ \\
\hline
\hline
\label{table:data}
\end{tabular}
\\$^{(a)}$ Minimum fraction of shock powered placed into non-thermal particles (eq.~[\ref{eq:entmin}]),\\
 calculated assuming radiative shocks and (conservatively) $\epsilon_{\gamma} = 0.2$ ($\S\ref{sec:hadronic}, \ref{sec:leptonic}$).\\ $^{(b)}$ Fraction of optical luminosity powered by shocks (eq.~[\ref{eq:fsh}]),\\ calculated assuming radiative shocks and (conservatively) $\epsilon_{\gamma} = 0.2$.
\end{minipage}
\end{table}

\section{Discussion and Conclusions}
\label{sec:discussion}

The high densities within the nova ejecta at early times coincident with the observed gamma-rays imply that the shocks responsible for this emission are likely to be radiative ($\S\ref{sec:radiative}$).  Though not observed directly, thermal X-rays from the shocks are absorbed by the ejecta and reprocessed to optical frequencies ($\S\ref{sec:optical}$).\footnote{Nova Del did show X-ray emission at later times, after the epoch of peak gamma-ray and optical emission.  This presumably occurred after the ejecta had expanded or became sufficiently photo-ionized to reduce the column of neutral absorbing gas (see below).}  Based on the observed values of $L_{\gamma}/L_{\rm opt}$, for the highest gamma-ray efficiencies $\epsilon_{\rm nth} \sim 0.1$ expected in hadronic scenarios, we conclude that a significant fraction $f_{\rm sh} \gtrsim 10$ per cent and $\gtrsim 1$ per cent of nova optical emission in nova Sco and nova Del, respectively, is powered by shocks.  These limits approach unity for more realistic efficiencies $\epsilon_{\rm nth} \lesssim 0.01$ expected from the difficulty of accelerating protons given the expected geometry of the magnetic field in the pre-shocked ejecta (see below).  

This shock-powered emission mechanism is similar to that at work in some core collapse supernovae when the ejecta from the exploding star interacts with dense circumstellar matter (e.g.~\citealt{Chevalier&Fransson94}; \citealt{Larsson+11}), which in extreme cases powers the most luminous supernovae yet discovered (\citealt{Smith+07}).  Our finding that a large fraction of nova optical emission is powered by shocks requires a revision of standard models that instead assume nova thermal emission results from the direct outwards transport of thermal energy released in the white dwarf envelope by nuclear burning (e.g.~\citealt{Hillman+14}).  The ultimate energy source (the thermonuclear runaway) is not in question, but shock-powered emission taps into the differential kinetic energy of the outflow instead of its thermal content advected or diffused from small radii.  As suggested by \citet{Metzger+14}, shock-powered optical emission might also help explain some of the irregular light curve shapes, including plateaus and secondary maxima, observed in some novae (e.g., \citealt{Strope+10}).
A similar mechanism of internal shocks within a time-variable outflow may power an optical transient from the remnant produced by the merger of binary white dwarfs (\citealt{Beloborodov14}).     

If the majority of nova optical light is powered by radiative shocks, then the ratio of optical and gamma-ray fluxes should be relatively constant in time, assuming that the microphysical parameters of the shocks remain constant (and that the conditions $\tau_{\rm p-p} > v_{4}/c$ and $t_{\rm cool}/t < 1$ remain satisfied in hadronic and leptonic scenarios, respectively).  Nova Sco is consistent with a temporally constant ratio $L_{\gamma}/L_{\rm opt}$, while nova Del shows evidence for the gamma-ray emission becoming relatively stronger with time (Fig.~\ref{fig:ratio}).  The apparent increase of $L_{\gamma}/L_{\rm opt}$ in nova Del could result from the reverse shock thermal emission becoming less radiatively efficient with time, i.e. $\chi \propto t$ (eq.~[\ref{eq:cool}]).  Alternatively, it may reflect the early transition in this event to a nebular phase (as occurred on Day 11; \citealt{Skopal:2014p3414}), resulting in a loss of flux out of the measured optical band and into the UV.  If nova shocks are indeed radiative, then we predict that $L_{\gamma}/L_{\rm opt}$ should not exceed $\sim 0.06$ at any time, since this represents the limit of shock-dominated emission $f_{\rm sh} = 1$ (eq.~[\ref{eq:fsh}]) for realistic maximum efficiencies $\epsilon_{\gamma} \lesssim 0.3$, $\epsilon_{\rm nth} \lesssim 0.2$.

The gamma-ray to optical flux ratio also places a lower limit on the acceleration efficiency $\epsilon_{\rm nth}$ of relativistic non-thermal particles at non-relativistic shocks.  For nova Sco and nova Del we constrain $\epsilon_{\rm nth} \gtrsim 10^{-2}$ and $10^{-3}$, respectively (Table \ref{table:data}).  Unlike traditional studies modeling particle acceleration in supernova remnants, nova shocks evolve in real-time, in principle allowing for the study of time-dependent effects.  Models used to constrain particle acceleration in supernova remnants also sometimes require assumptions about the escape fraction of relativistic particles, e.g. if gamma-rays are produced by the collision of relativistic protons with nearby molecular clouds of irregular geometry.  By contrast, nova ejecta serves as a relatively efficient hadronic ``calorimeter" since relativistic protons are probably trapped when gamma-rays are observed (eq.~[\ref{eq:tpp}]).  

Neither hadronic nor leptonic scenarios for nova gamma-ray emission can be ruled out by modeling of the $\gamma-$ray spectrum alone (\citealt{Ackermann+14}).  However, the tension between the value $\epsilon_{\rm nth} \gtrsim 10^{-2}$ we find is required in nova Sco and the much lower electron acceleration efficiency $\epsilon_{\rm nth} \lesssim 10^{-3}$ inferred from observations of supernova remnants (e.g.~\citealt{Morlino&Caprioli12}) and theoretical modeling (\citealt{Kato14}; \citealt{Park+14}) appears to disfavor the leptonic scenario.  Leptonic models also require a flat injected spectrum $p = 2$ to be energetically feasible (eq.~[\ref{eq:egammae}]).  In contrast, ion acceleration efficiencies as high as $\epsilon_{\rm nth} \sim 0.1$ are found for non-relativistic shocks by hybrid PIC simulations, but only in cases when the upstream magnetic field is parallel to the shock normal (\citealt{Caprioli&Spitkovsky14}); for perpendicular fields essentially no ion acceleration is seen, also inconsistent with our lower limits on $\epsilon_{\rm nth}$.  Hadronic scenarios involving ion acceleration at quasi-parallel shocks thus appear to be the most viable source of relativistic particles responsible for nova gamma-rays.  

However, this conclusion is problematic.  Most of the power dissipated in the system occurs at the reverse shock, making it the most natural location for particle acceleration.  If the fast outflow is ejected over a timescale which is long compared to the dynamical time near the surface of the hydrostatic white dwarf envelope (i.e., a quasi-steady wind), then its magnetic field will become dominated by its azimuthal component at the much larger radii where the reverse shock occurs ($\S\ref{sec:hadronic}$).  Such a field, being perpendicular to the radial shock normal, is not conducive to ion acceleration.  

This mystery could be resolved if particle acceleration occurs at localized regions of quasi-parallel shocks, creating special regions of efficient acceleration  ($\epsilon_{\rm nth} \sim 0.1$) with $\epsilon_{\rm nth} = 0$ across the larger bulk, thereby resulting in an {\it average} value of $\epsilon_{\rm nth} \sim 10^{-3}-10^{-2}$ consistent with observations.  These localized regions of conducive field geometry could be caused by global asymmetries in the ejecta, such as oblique shocks occurring where the edge of the slow ejecta torus meets the faster bipolar wind (Fig.~\ref{fig:schematic}).  Alternatively, inhomogeneity of the nova ejecta (``clumpiness") could also result in localized regions of oblique shocks between the clumps.  Indeed, low ejecta filling factors are inferred observationally by emission line modeling of the nebular phase (e.g.,~\citealt{Shore+13}).  Radial fields may also be produced near the shock as the result of Rayleigh-Taylor instabilities at the interface between the fast now outflow and the dense central shell (as may occur in supernova remnants; e.g., \citealt{Blondin&Ellison01}).

Leptonic versus hadronic models for nova gamma-ray emission may be further distinguished by their predictions for non-thermal X-ray emission.  In nova Sco, {\it Swift} XRT observations placed an upper limit of $F_{\rm X} < 9\times 10^{-14}$ erg s$^{-1}$ cm$^{-2}$ on the 0.3$-$10 keV flux on days 22$-$41 \citep{Page:2012p3257}, i.e. three orders of magnitude less than the simultaneous LAT-detected flux of $F_{\gamma} \sim 2\times 10^{-10}$ erg cm$^{-2}$ s$^{-1}$.  Such a low value of $F_{\rm X}/F_{\gamma}$ would at first appear to be barely compatible with the leptonic scenario, even when considering the most conservative case of $p = 2$ and a slow cooling spectrum $\nu F_{\nu} \propto \nu^{1/2}$ from keV to $\sim$ GeV energies (eq.~[\ref{eq:dEgammae}]).  The hadronic scenario fares even worse: secondary $e^{\pm}$ pairs from $\pi^{\pm}$ decay carry a similar total energy to that released in $\gamma-$rays from $\pi_0$ decay.  These pairs' energies, $\sim 0.1-1$ GeV, correspond to Lorentz factors $\gamma_e \sim 100-1000$ that will upscatter $\sim$ eV optical to energies $\sim 10-1000$ keV with reasonably high efficiency (eq.~[\ref{eq:tcool2}]).  However, a more detailed calculation must account for the important role of coulomb losses on the energy of the X-ray emitting electrons, which are likely severe due to the high densities of nova shocks; this will likely suppress the X-ray emission below that obtained by a naive extension of the IC or bremsstrahllung spectrum to lower energies.     

This dearth of early $\sim$ keV emission from nova shocks could result from the high column of neutral, X-ray absorbing gas ahead of the shock at times coincident with the gamma-ray emission (\citealt{Metzger+14}).  However, higher energy X-rays $\gtrsim 10-100$ keV could escape without photoelectric attenuation\footnote{Although attenuation by inelastic electron scattering could be important for sufficiently high optical depths, leading to hard X-ray suppression.}, motivating one to consider the prospects for nova detection with NuSTAR (\citealt{Harrison+13}), which has unprecedented X-ray sensitivity above 10 keV.  If a fraction $f_{X}$ of the LAT luminosity is radiated as X-rays of energy $\epsilon_{X} \sim 30$ keV, this results in a number flux of $\dot{N}_{\rm X} \sim f_{X}F_{\gamma}A_{\rm eff}/\epsilon_{X} \sim 4\times 10^{-4}(f_X/0.01)$ s$^{-1}$ for a typical value $F_{\gamma} \sim 10^{-10}$ erg cm$^{-2}$ s$^{-1}$ (\citealt{Ackermann+14}) and the NuSTAR effective area $A_{\rm eff} \approx 200$ cm$^{2}$ (\citealt{Harrison+13}; their Fig.~2).  Given the NuSTAR estimated background of $\dot{N}_{\sigma} \sim 2\times 10^{-3}$ counts s$^{-1}$ in this energy range across the point spread function (\citealt{Harrison+13}; their Fig.~11), we estimate that the integration time required for a 3$\sigma$ detection is approximately $9\dot{N}_{\sigma}/\dot{N}_{\rm X}^{2} \sim 100(f_{X}/0.01)^{-2}$ ks.  Reasonable constraints on $f_{X} \sim 0.01$ could thus be achieved for a modest 100 ks exposure.  The hard X-ray imager on ASTRO-H should also prove useful for nova follow-up.

\citet{Orio+14} reported NuSTAR observations of the fast symbiotic nova V745 Sco within 10 days of the outburst, placing an upper limit of $\lesssim 10^{-11}$ ergs cm$^{-2}$ s$^{-1}$ on the $\gtrsim 50$ keV emission.  Gamma-ray emission was detected by LAT at $2-3\sigma$ significance and a flux of $\sim 10^{-10}$ erg cm$^{-2}$ s$^{-1}$, but lasting only for 2 days near the onset of the outburst (\citealt{Cheung+14}), approximately a week before the NuSTAR observations and when (perhaps not coincidentally in light of our results on shock-powered optical emission) the optical light curve was at its maximum.  By the time of the NuSTAR observations, the optical flux had decreased by a factor of $\sim$100 from its peak value, such that if $F_{\gamma}/F_{\rm opt}$ had remained approximately constant in time (as in the novae in our sample; Fig.~\ref{fig:ratio}), then the NuSTAR upper limit correponds to $f_{X} \lesssim 10$ in the notation above, i.e. not particularly constraining.  We strongly encourage future NuSTAR observations of gamma-ray novae, ideally closer to the optical peak and coincident with LAT detections.

In this paper we have focused on internal shocks in classical novae instead of symbiotic novae, because in the symbiotic case shocks and particle acceleration can result also from the interaction between the nova ejecta and the effectively stationary dense wind of the red giant companion (\citealt{Abdo+10}, \citealt{Martin&Dubus13}).  Our analysis relies on the assumption of radiative shocks internal to the nova ejecta, as is justified by the higher mass outflow rates $\dot{M} \sim 10^{-3}M_{\odot}$ yr$^{-1}$ and densities that characterize nova ejecta as compared to the lower values $\lesssim 10^{-5}M_{\odot}$ yr$^{-1}$ characteristic of red giant winds.  Nevertheless, much of the same physics of relativistic particle acceleration could apply to both scenarios.  Likewise, the presence of a dense equatorial structure associated with the nova ejecta could also be present in symbiotic systems (as was inferred in V407 Cyg; \citealt{Martin&Dubus13}).

An additional prediction of the favored hadronic model is a GeV neutrino flux comparable to the gamma-ray flux that may be detected by future experiments (\citealt{Razzaque+10}).  Depending on the maximum ion energy, novae may be important sources of TeV gamma-ray emission.  \citet{Sitarek+15} recently presented upper limits on TeV emission from V407 Cyg by the MAGIC telescope (\citealt{Aliu+12}) which lie a factor of $\sim 10$ below the extrapolated Fermi LAT spectrum; similar TeV upper limits on V407 Cyg were reported by VERITAS (\citealt{Aliu+12}).  The observed cut-off in the photon spectrum may be due to an intrinsic cut-off in the accelerated ion spectrum (eqs.~[\ref{eq:Emax}, \ref{eq:Emax2}]), or it may result from $\gamma-\gamma$ absorption due to $e^{\pm}$ pair-creation attenuation by the nova optical light.  TeV photons can pair create off target photons of energy $E_{\rm opt} \sim 1$ eV, resulting in an optical depth
\begin{eqnarray}
\tau_{\gamma-\gamma} \approx n_{\gamma}\sigma_{\gamma-\gamma}R_{\rm ej} \approx 0.3 \tau_{\rm opt} \left(\frac{E_{\rm opt}}{\rm eV}\right)^{-1}\left(\frac{L_{\rm opt}}{10^{38}\,{\rm erg\,s^{-1}}}\right)v_{8}^{-1}t_{\rm wk}^{-1}
\label{eq:TeV}
\end{eqnarray}
near unity, where $\sigma_{\gamma-\gamma} \sim 10^{-25}$ cm$^{2}$ is the photon-photon absorption cross section near threshold and $n_{\gamma} = L_{\rm opt}\tau_{\rm opt}/(4\pi c R_{\rm ej}^{2}E_{\rm opt})$ is the energy density of the target optical photons near the shock  (for $\tau_{\rm opt} > 1$).  We strongly encourage future additional TeV follow-up of novae near and after optical peak, even if not first detected by {\it Fermi}, as detections or upper limits can be used to place meaningful constraints on the maximum accelerated particle energy and the location of the gamma-ray emission.

The sample of gamma-ray novae should expand rapidly in the next few years thanks to anticipated enhancements in the sensitivity of {\it Fermi} LAT resulting from improvements in its ability to perform low-level event reconstruction.  The method for jointly analyzing optical and gamma-ray data employed in this paper will thus soon be applicable to a greater sample of events.  Our results highlight the importance of obtaining broad (ideally {\it bolometric}) frequency coverage of nova light curves, including near infrared and ultraviolet wavelengths, at times coincident with the gamma-ray emission.   The utility of our model rests heavily on the assumption of radiative shocks, which depends sensitively on the quantity of mass and velocity of the nova ejecta, and how it is partitioned between the fast and slow components.  Additional radio monitoring of gamma-ray novae is thus also key to better constraining the ejecta mass from these systems and its geometry, thereby strengthening or refuting the argument for radiative shocks.  

\section*{Acknowledgments}
BDM acknowledges helpful conversations with Damiano Caprioli, Koji Mukai, Jeno Sokoloski, Andrey Vlasov, and Bob Williams.  We thank Fred Walter for providing optical data on nova Sco.  BDM gratefully acknowledges support from NASA {\it Fermi} grant NNX14AQ68G, NSF grant AST-1410950, and the Alfred P. Sloan Foundation.  We thank the reviewer, Guillaume Dubus, for helpful criticism and suggestions that led to an improvement of the manuscript.  

\bibliographystyle{mn2e}

\end{document}